# Raman Spectroscopy of Single Nanoparticles in a Double-Nanohole Optical Tweezer System

Steven Jones, Ahmed A. Al Balushi, Reuven Gordon*
Department of Electrical and Computer Engineering,
University of Victoria, Victoria, BC, Canada V8P5C2
*rgordon@uvic.ca

**ABSTRACT**

A double nanohole in a metal film was used to trap nanoparticles (20 nm diameter) and simultaneously record their Raman spectrum using the trapping laser as the excitation source. This allowed for the identification of characteristic Stokes lines for titania and polystyrene nanoparticles, showing the capability for material identification of nanoparticles once trapped. Increased Raman signal is observed for the trapping of multiple nanoparticles. This system combines the benefits of nanoparticle isolation and manipulation with unique identification.

The first demonstrations of gradient force optical tweezers recognized the limitation of trapping objects smaller than 100 nm due to the requirement for large optical intensities [1]. Many nanophotonic and nanoplasmonic approaches have been used to achieve trapping of nanoparticles with subwavelength confined optical fields [2-10]. In 2009, a circular nanohole in a metal film was used to trap 50 nm dielectric particles with low optical power [11]. That work used the strong influence of the aperture's transmission due to dielectric loading to achieve efficient trapping [12].

Several works have developed shaped nanoholes to achieve trapping of even smaller nanoparticles [13-17]. Double nanoholes (DNHs) that strongly confine the electromagnetic field at the cusps where the holes overlap have been used for trapping (and unfolding) a single protein [14], a 12 nm silica sphere [15] and a single DNA molecule [16]. Rectangular nanoholes have trapped 22 nm polystyrene particles [18]. Similar to the DNH, the bowtie nanohole has been fabricated on the end of a tapered optical fiber, allowing for not only the trapping but also the translation of 50 nm fluorescent beads [17]. Apertures for trapping have also been integrated on other near-field fiber probes [19] as well as cleaved optical fibers [20]. A recent work has used the bowtie nanoholes to isolate quantum dots and excite them with two-photon luminescence [21].

While these works illustrate the potential of nanoholes to isolate nanoparticles and manipulate them, it is of considerable interest to identify the trapped nanoparticle. Recently, we have developed an approach to excite the vibrational modes of the trapped nanoparticles, proteins and DNA in the low-wavenumber regime [22,23]. This requires multiple trapping lasers and so far has only been demonstrated for extremely low frequency vibrational modes that probe the nanoparticle size, shape and mechanical properties. To achieve conventional Raman spectra for the identification of the trapped nanoparticle material, here we integrate a Raman spectroscopy setup with our DNH tweezer system. A recent report showed multiple polystyrene nanoparticles (of the order of 100) detected with Raman using a nanohole [24]. Here we demonstrate single nanoparticle sensitivity to achieve characteristic Raman peaks for titania and polystyrene nanoparticles. This combines surface enhanced Raman spectroscopy, which has enabled single molecule sensitivity [25-27], with nanohole optical tweezers for the isolation and manipulation of nanoparticles.

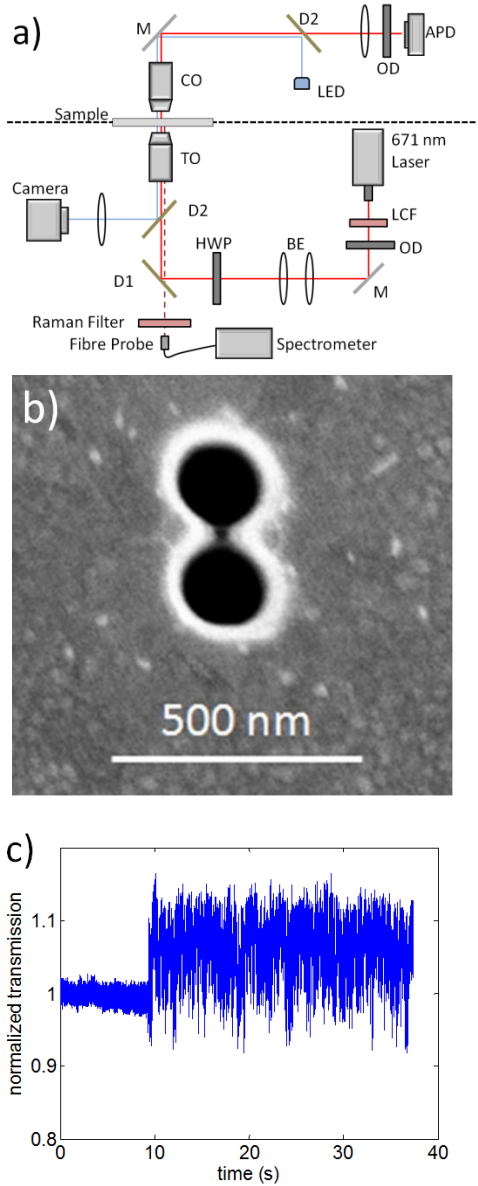

Figure 1 (a) Schematic of the trapping setup used to obtain single nanoparticle Raman spectra. APD - avalanche photodiode; BE - beam expander; CO - 10x condenser objective; D1 - 685 nm long pass dichroic; D2 - 650 nm long pass dichroic; HWP - half wave plate; LCF - laser clean-up filter; M - silvered mirror; N1 – laser notch filter; OD – optical density filter; TO – 50x trapping objective. (b) Scanning electron microscope image of the dual nanohole aperture used in trapping. (c) Characteristic single particle trapping event (polystyrene).

Figure 1(a) shows a schematic of the optical tweezer system with integrated Raman detection. While similar to our past works there are notable differences in this setup: it uses a 671 nm solid state laser (100 mW, reduced to around 10 mW at the nanoaperture due to filters) with a notch filter (Semrock, LL01-671) and a high-pass dichroic (Semrock, FF685-Di02) to isolate the Rayleigh line from the detected Raman spectra. The Raman spectrum was recorded in the reflection path by collecting with a broad area optical fibre and a cooled CCD spectrometer (Ocean Optics, QE Pro).

Figure 1(b) shows a scanning electron microscope image of typical double nanoholes used in the setup. The gap sizes were slightly larger than 20 nm to allow for trapping of 20 nm particles without steric hindrance (although gaps down below 10 nm have been achieved in the past [28]). Figure 1(c) shows the transmission through the aperture with a characteristic step increase when a particle is trapped in the aperture. The transmission is normalized to the mean value before trapping. There is also an increase in the fluctuations of the transmitted power which is due to the motion of the particle in the trap. We have used this in the past to characterize the strength of the optical tweezer [29] as well as identify the size of proteins [30].

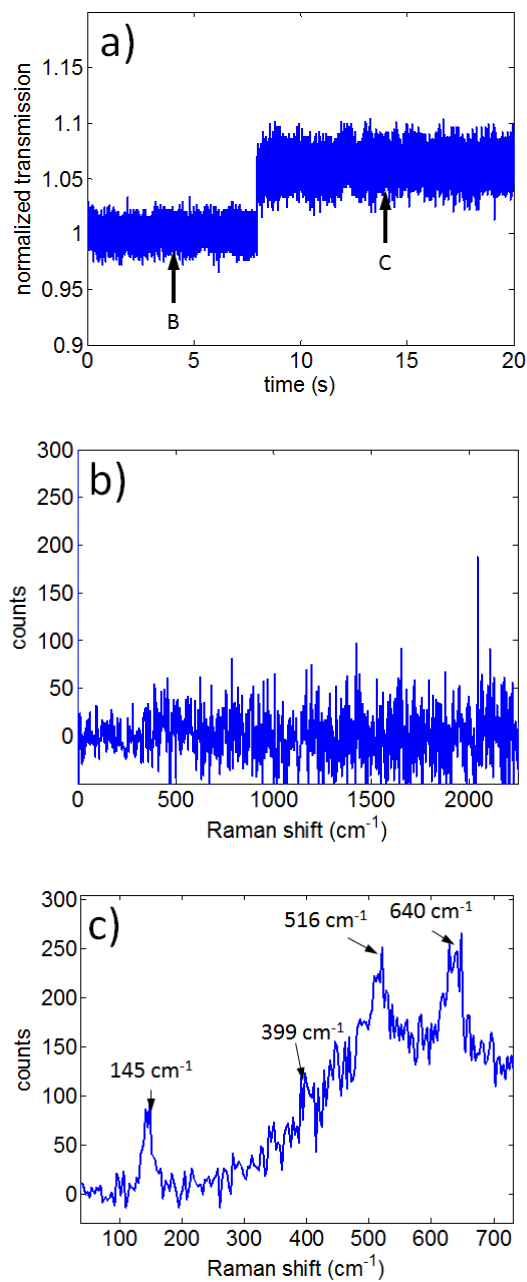

Figure 2 (a) Trapping event of 20 nm titania used for obtaining Raman spectrum, B is the untrapped state and C is the trapped state. The Raman spectra in the (b) untrapped and (c) trapped states.

Figure 2(a) shows a trapping event for a 21 nm titania nanoparticles (Sigma: 718467-100G). Prior to trapping, the Raman spectrum (background subtracted) shows no peaks, as seen in Figure 2(b). After trapping, there is a Stokes' peak at 145, 399, 516 and 640 cm$^{-1}$ (Figure 2(c)), which agree well with previously reported values for titania nanoparticles

[31]. The attenuation of Raman peaks from 0 to 500 cm$^{-1}$ is expected to be due to the band edge of the dichroic filter. There also appears to be a fluorescence background from the titania nanoparticle.

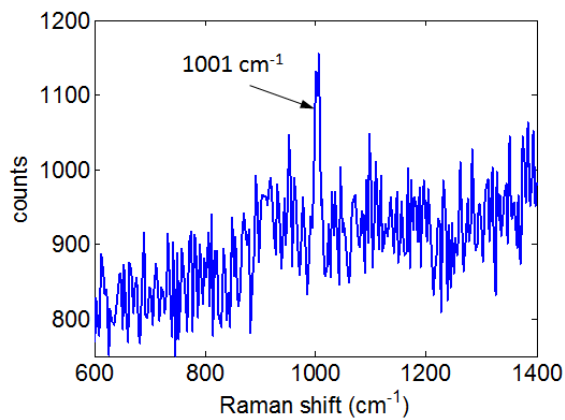

Figure 3 Raman spectra of trapped 20 nm polystyrene particle (5 min integration time).

To show that this configuration can be used to identify different nanoparticles, we repeated the experiment for 20 nm polystyrene spheres (Thermo Scientific, 3020A). The trapping event for polystyrene is shown in Figure 1(c). The corresponding Raman spectrum is shown in Figure 3. The main peak for polystyrene at 1001 cm$^{-1}$ is clearly seen. Additional Stokes peaks were too small to be observed, which is expected for polystyrene (the next largest peak in the detection range is around 4 times smaller).

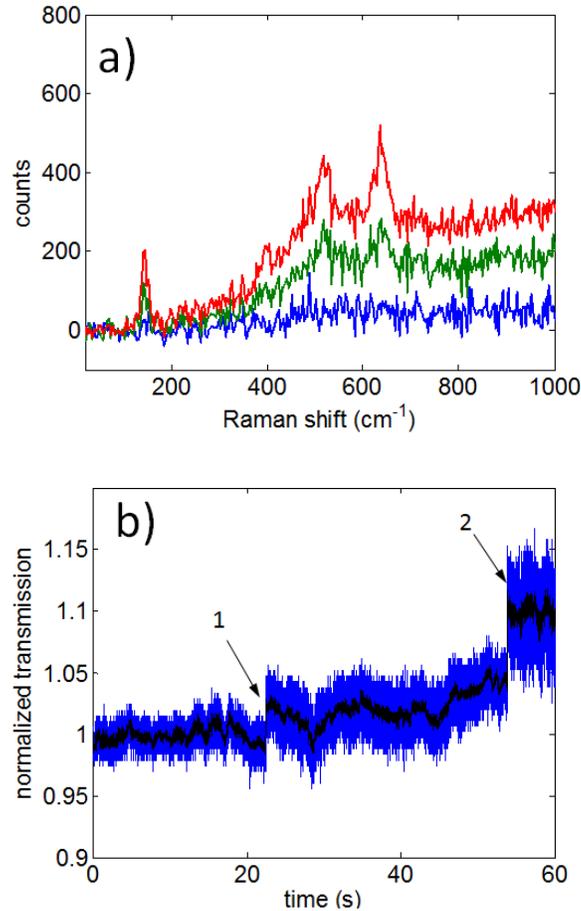

Figure 4 (a) Raman spectra of titania nanoparticles for multiple trapping events. (b) Time series illustrating trapping events 1 and 2, the black line is a filtered time series to better illustrate the stepped transmission increases at discrete times (in contrast to slower drift variations).

In some cases, trapping of multiple particles was observed by multiple steps in the transmission spectrum (Figure 4(a)). In these cases, the Raman spectrum also showed an increase, approximately doubling for the second particle. Repeated measurements showed the same results for both the titania and polystyrene nanoparticles (in particular, the titania results have been reproduced for more than 20 trapping events).

DNHs also have the advantage of rapidly dissipating heat away from the nanohole. It has been predicted that the heating of nanoholes is over three orders of magnitude lower than a corresponding nanoparticle (for the same local field intensity) due to the high thermal conductivity of the surrounding gold film [32]. In the future, it may be possible to quantify the heating by comparing Stokes and anti-Stokes lines. This will require modifying the setup with a bandpass filter that is not possible in the current configuration.

In conclusion, we have demonstrated the ability to record the Raman spectra of individual and multiple nanoparticles in a DNH laser tweezer system. This allows for identifying the nanoparticle trapped for further investigation. This work may be used to study the characteristics of nanoparticles already investigated with the nanohole tweezers, such as quantum dots, proteins, DNA, and viruses. Such identification capability will allow operation in heterogeneous solutions. The Raman spectra also allows the potential of analyzing material changes due to interactions (e.g., chemical reactions, binding, strain), as well as probe the local temperature in the aperture.

**Acknowledgements:**

SJ acknowledges support from a UVic Graduate Fellowship. This work is supported by an NSERC Discovery Grant.